\documentclass{eas}
\usepackage{graphicx}
\begin{document}

\title{Evolution of the Milky Way Halo by accretion of dwarf satellite galaxies} 
\thanks{This work was supported by the Initiative Kolleg (IK) 'The Cosmic Matter Circuit' I033-N program
and computational time on the Vienna Scientific, Grape \& Astro Clusters.}
\author{Mykola Petrov}\address{Institute of Astronomy,
University of Vienna, Austria;
  \email{mykola.petrov@univie.ac.at} 
\& \email{gerhard.hensler@univie.ac.at}}
\author{Gerhard Hensler}\sameaddress{1}
\begin{abstract}
Within the Cold Dark Matter scenario the hierarchical merging paradigm
is the natural result to form massive galactic halos by the minor 
mergers of sub-halos and, by this, inherently their stellar halo. 
Although this must be also invoked for the Milky Way, the context of 
chemical and kinematic coherence of halo stars and dwarf spheroidal
galaxies is yet unsolved a focus of present-day research. 
To examine this issue we model the chemo-dynamical evolution of the 
system of satellites selected from the cosmological Via Lactea II
simulations to be similar for the Milky Way environment but at an 
early epoch.
\end{abstract}
\maketitle
\section{Introduction}
From the hierarchical merging paradigm and from numerical
simulations of $\Lambda$CDM cosmology numerous Dark Matter 
subhalos are expected to assemble around a massive halo and
to accumulate its mass. If they have already experienced star
formation (SF), their stars should be merged into a spheroid and 
be identifyable by their kinematics and chemical abundances. 
Although these low-mass sub-halos and their baryonic content as
dwarf spheroidal galaxies (dSphs) therefore serve as the 
key to pinpoint this cosmological $\Lambda$CDM description,
observational detections of stellar streams within the Milky Way 
(MW) halo are rare though requested from accretion models 
(\cite{Johnston08}), the stellar abundances in present-day dSphs 
deviate clearly from the halo (\cite{Tolstoy09}), and evolutionary 
models of dSph SF, chemistry, and gas expulsion are still too 
simplistic. 
Not only that their accretion epoch occurred continuously over 
the Hubble time instead of a short event (\cite{Prantzos08}), and 
their gas is not only removed by tidal and ram-pressure stripping 
(\cite{Mayer07}) but is instead re-accreted while orbiting the MW 
(\cite{Bouchard09}), also their consideration as isolated systems 
lacks reality (\cite{Revaz09}).

\section{Modeling}
\label{sec:1}
To model the evolution of the system of dSphs in the gravitational
field of the MW for which the accretion by the host galaxy is probable
within the Hubble time, with the highest possible mass resolution 
the Via Lactea II (\cite{Diemand}) simulation was used to follow 
250 sub-haloes as DM progenitors of dSphs in the DM mass range of
$10^6 M_{\odot} < M_{sat} < 6\cdot10^8 M_{\odot}$ from $z=4.56$ for
the first Gyr. Because the production and chemical evolution of 
11 elements is intended to be treated within a proper computational 
time the number of gas particles is limited to $10^6$ as also the DM.
These facts limit the radius of consideration to within a radius of 
40 kpc of the MW center of mass.
For the simulations an advanced version of the single-gas 
chemo-dynamical SPH/N-body code (\cite{PH10}) is applied.

\begin{figure}
\begin{tabular}{ll}
\resizebox{0.48\columnwidth}{!}{%
  \includegraphics{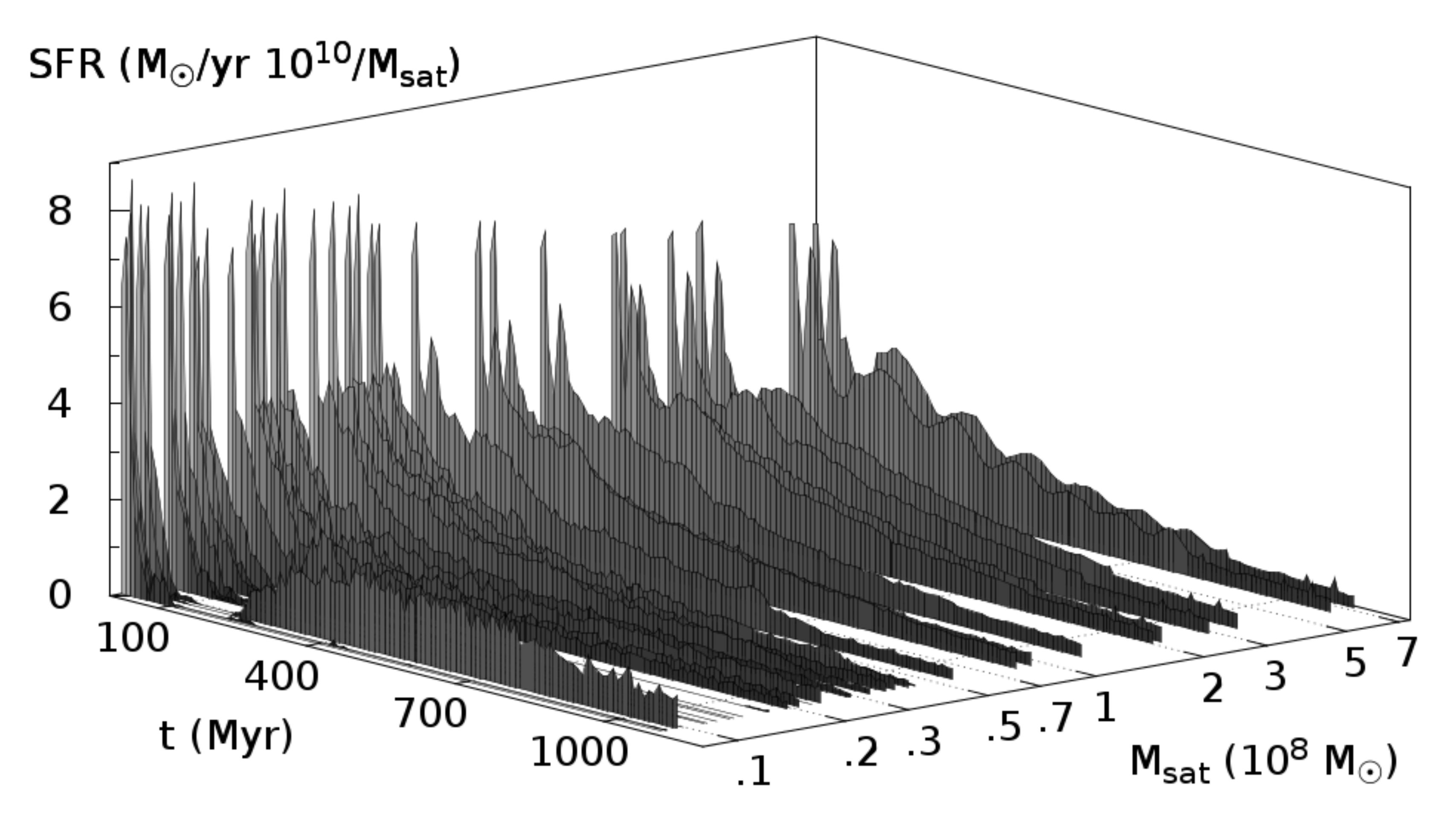} }
&
\resizebox{0.48\columnwidth}{!}{%
  \includegraphics{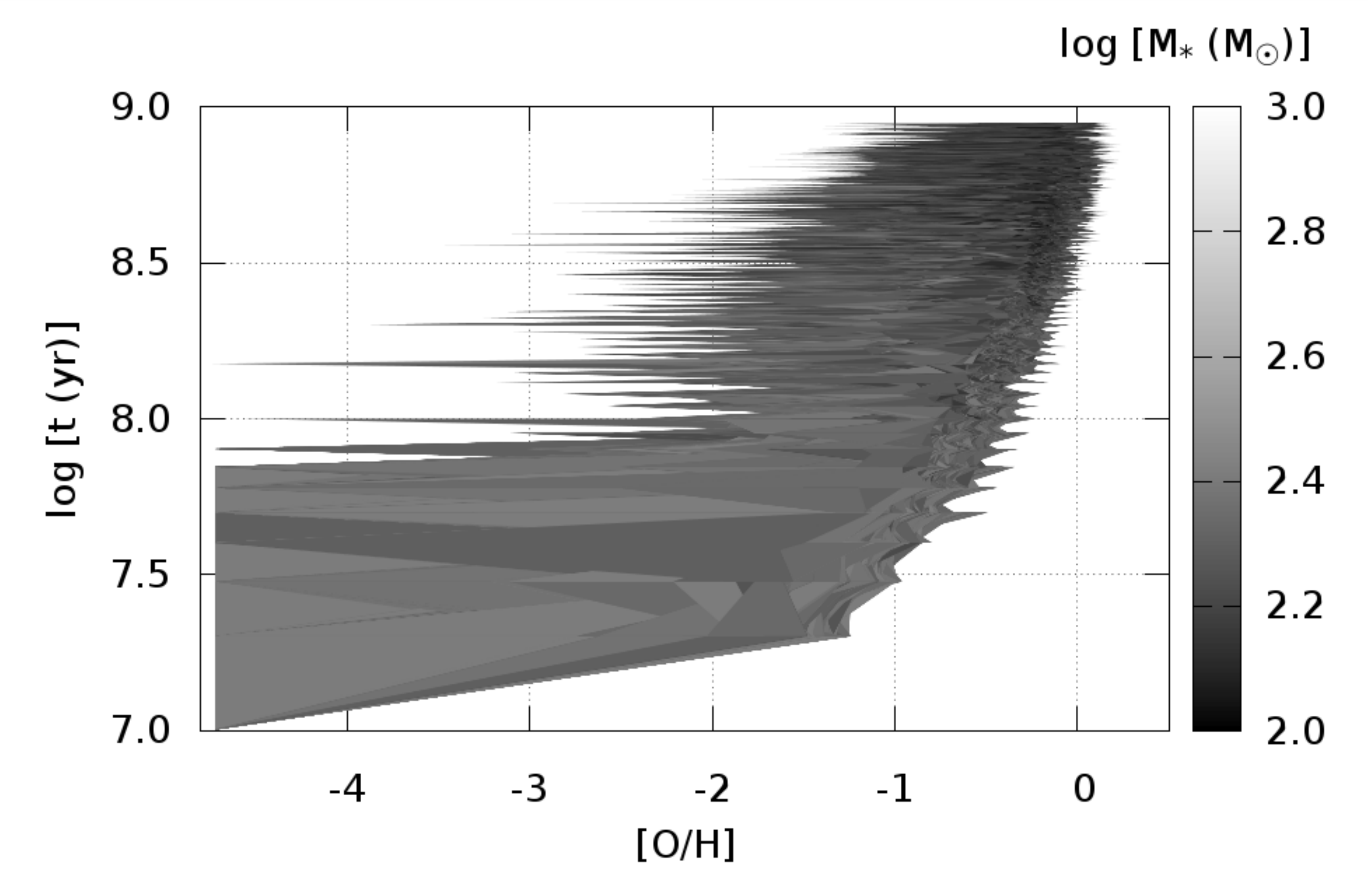} }
\end{tabular}
\caption{{\it Left:} SFR of the first $1Gyr$ as function of time and total mass of satellites.
{\it Right:} Formation epoch $t$ of stars with stellar oxygen 
abundance, $[O/H]$, colour-coded for the newly formed stellar mass 
(on logarithmic scales).}
\label{fig:1}
\end{figure}

Starting in virial equilibrium of a $10^4$ K warm gas, because 
re-ionization is improbable to have affected the Local Group dSphs
(\cite{Grebel04}), cooling allows the gas particles to achieve SF conditions 
in all satellites (Fig.~\ref{fig:1}), but its efficiency 
directly depends on the mass of a satellite and its dynamical history 
(merging with other satellites or disruption by the MW potential). 
The stellar feedback by supernovae type II (SNeII) releases sufficient 
energy to expel hot gas from the main bodies of less massive dSphs facilitated 
by tidal interactions. This gas accumulates in the MW halo 
while massive dSphs merge an continue SF.
In Fig.~\ref{fig:1} (right panel) SF epoch is shown as function 
of the oxygen abundance of newly formed stars. 
For the first $10^8$ yr of the simulation there is a considerable
variance of stellar oxygen abundance in the whole system 
$(-5 \leq [O/H] \leq -0.5)$ reflecting the very inhomogeneous 
production and distribution of enriched gas.
After $10^8$ yr the merger of sub-satellites ISM promotes the mixing 
of heavy elements. Finally, almost complete recycling of the gas 
erases the abundance inhomogeneities so that O in stars converges 
to $-1 \leq [O/H] \leq 0$ with a small dispersion.

In a comprehensive paper (\cite{PH10}) we will present detailed 
analyseds of the SF history, gas exchange, stellar abundance evolution of dSphs and the MW halo in the early universe and will discuss their 
implications for our cosmological picture. 

%


\begin{thebibliography}{99}
\vspace{-0.2cm}
\bibitem[Bouchard \etal\  2009]{Bouchard09}
Bouchard, A. \etal\  2009, AJ, 137, 3038
\bibitem[Diemand \etal\ 2008]{Diemand}
Diemand, J. \etal\ 2008, Nature, 454, 735
\bibitem[Grebel \etal\ 2004]{Grebel04}
Grebel, E.K. \& Gallagher, J.S. 2004, ApJ, 610, L89
\bibitem[Johnston \etal\  2008]{Johnston08}
Johnston, K. \etal\  2008, ApJ, 689, 936
\bibitem[Mayer \etal\ 2007]{Mayer07}
Mayer, L. \etal\ 2007, Nature, 445, 738
\bibitem[Petrov \& Hensler 2010]{PH10}
Petrov, M. \& Hensler, G. 2010, ApJ, submitted
\bibitem[Prantzos (2008)]{Prantzos08}
Prantzos, N. 2008, A\&A, 489, 525
\bibitem[Revaz \etal\ 2009]{Revaz09}
 Revaz, Y. \etal\ 2009, A\&A, 501, 189
\bibitem[Tolstoy \etal\  2009]{Tolstoy09}
Tolstoy, E., \etal\  2009, ARA\&A, 47, 371

\end{thebibliography}
\end{document}